# A Localization Method for the Internet of Things

Zhikui Chen · Feng Xia · Tao Huang · Fanyu Bu · Haozhe Wang

**Abstract:** Many localization algorithms and systems have been developed by means of wireless sensor networks for both indoor and outdoor environments. To achieve higher localization accuracy, extra hardware equipments are utilized by most of the existing localization solutions, which increase the cost and considerably limit the location-based applications. The Internet of Things (IOT) integrates many technologies, such as Internet, Zigbee, Bluetooth, infrared, WiFi, GPRS, 3G, etc, which can enable different ways to obtain the location information of various objects. Location-based service is a primary service of the IOT, while localization accuracy is a key issue. In this paper, a higher accuracy localization scheme is proposed which can effectively satisfy diverse requirements for many indoor and outdoor location services. The proposed scheme composes of two phases: 1) partition phase, in which the target region is split into small grids; 2) localization refinement phase, in which a higher accuracy of localization can be obtained by applying an algorithm designed in the paper. A trial system is set up to verify correctness of the proposed scheme and furthermore to illustrate its feasibility and availability. The experimental results show that the proposed scheme can improve the localization accuracy.

Z.K. Chen
School of Software, Dalian University of Technology, Dalian 116620, P.R. China
e-mail: zkchen@dlut.edu.cn

F. Xia (Corresponding author)
School of Software, Dalian University of Technology, Dalian 116620, P.R. China
e-mail: f.xia@ieee.org

T. Huang
School of Software, Dalian University of Technology, Dalian 116620, P.R. China
e-mail: ht2411@gmail.com

F.Y. Bu
School of Software, Dalian University of Technology, Dalian 116620, P.R. China
e-mail: pfy@imfec.edu.cn

H.Z Wang
School of Software, Dalian University of Technology, Dalian 116620, P.R. China
e-mail: wang-haozhe@hotmail.com

# 1 Introduction

Computer to computer or machine to machine is interconnected via the ongoing Internet from 1990s. By this way, various resources can be shared with each other. Person to person or person to machine can be interconnected by the mobile phones and smart devices over wideband wireless communication systems and internet, so that communications between persons are realized anywhere, anytime. It is comprehensible that the next step is to interconnect various objects. As a consequence, the "Internet of Things" (IOT) emerges. The IOT uses some sensing devices such as sensors, Radio Frequency Identification (RFID), infrared, Bluetooth, Global Positioning System (GPS), laser scanner, cameras and some communication protocols connecting all the objects to accomplish smart identification, object tracking, localization, management and scheduling.

IOT may be simply regarded as the extension of Internet, but they are quite different. The IOT has various independent subsystems which can operate on the existing infrastructure to obtain the status information of devices and also control them. Generally, the IOT has three layers: the bottom layer is sensing layer, which employs various sensors to acquire the information of surrounding and transfers them to the upper layer, namely network layer. The network layer is mainly used to implement the data transmission, consisting of a great diversity of short distance heterogeneous communication networks, such as Zigbee, Bluetooth, WiFi, and long distance communication networks such as GPRS, 3G and so on. Data fusions may carry out in local region to reduce the traffic. The application layer processes the data delivered from the lower layer and analyzes its meanings to control corresponding devices.

Wireless sensor network (WSN) [9] as a subpart of IOT has been regarded as a promising data acquisition tool, which has been gradually used in industrial filed as well as national defense field. As a cross-discipline, it integrates micro-electrical technique, short-range communication technique and embedded system technique into together. However, a big challenge has been faced as its stringent constraints in energy consumption, memory storage and communication bandwidth. In spite of these, as a supporting technique in WSNs, localization problem has been given a great concern, especially in data marker, geographic routing and data aggregation algorithm, where data are meaningless if there is no location information.

Traditional location technique such as GPS cannot be used in WSNs directly, as its costly requirement of sophisticated equipment and high energy consumption, which have greatly constrained the application scale of WSNs. In WSNs energy conservation has been considered as the core issue and thus the costs as well as size of node should be as small as possible, to apply into large-scale applications for a long time. To address this issue, many localization algorithms developed do not use GPS technique directly, but employ it as an assistance in some cases, and more efforts are made on the excavation of WSNs itself. Till now, most of the localization algorithms proposed in the literature can be broadly classified into two categories: range-based localization [12, 13, 14, 15, 16, 17] and range-free localization [18, 19, 20, 21, 22, 23]. The former technique usually needs extra hardware to accomplish ranging, and then utilizes some algorithm to calculate coordinates. The latter technique exploits the characteristics of network connectivity [18, 19, 20] (such as DV-hop [18], MDS-MAP [19], Amorphous [20]), proximity information [21, 22, 23] (such as APIT [21], Centroid [22], Convex [23]) to realize a tolerable position estimation. Although it cannot provide such high accuracy as range-based schemes often do, it holds a cost-effective advantage, which is an enormous attraction for large-scale system deployment.

The remaining part of this paper is organized as follows: Section 2 summaries related works. The proposed algorithm is described in Section 3 Section 4 demonstrates the experimental results. Finally, Section 5 concludes this paper.

# 2 Related works

The concept of IOT originally was proposed by professor Ashton of MIT Auto-ID in the research of RFID in 1999. At that time the research primarily focus on RFID, to gain the object information by browsing the Internet address or database entry to achieve recognition for objects. After that, the scope of IOT has been expanded to many areas such as environment surveillance, health care, smart home, logistics, and forest-fire prevention and so on. In 2005 ITU give an Internet reports named "The Internet of Things" [7]. In this reports some technologies for IOT as well as opportunity, challenges are pointed out. In 2009 the European community drafts out a report named "Internet of things - An action plan for Europe" [8], in which Europe insisted on that by adopting a proactive approach they can play a leading role and people can reap the benefits from it.

Many researchers have a great interest in IOT, and made many creative achievements. Atzori et al [1] have reviewed many integrated technologies and communication solutions, as well as several concrete applications, like touch upon transportation, healthcare, smart environments and personal, social domain. Castellani et al. [3] have proposed a practical architecture for IOT, which is used to provide environment monitoring and localization services. Silverajan and Harju [4] suggest a communication protocol design approach to construct network which can span many wireless radio networks of varying link-level characteristics. For IOT, a highly interoperable and lightweight event-based framework has been propounded. Bohli et al. [5] hold the opinion that IOT can offer new enhanced services and applications which will benefit for total society, and the economics and price issues should be well considered when providing a sensor-based service. Khoo [6] reviews some issues and technologies which are necessary to enable RFID technology successfully used in IOT applications. Huircán [24] and Cho [25] present examples of localization system using Zigbee protocol in WSNs which is also suitable for IOT.

In IOT, data acquisition and sensing usually using wireless sensor technology, thus concerning the localization issue we can also refer to WSNs in some extent. In WSNs, many localization algorithms have been developed, all of which can be roughly categorized into one of the follows: range-based localization [12, 13, 14, 15, 16, 17] and range-free localization [18, 19, 20, 21, 22, 23]. Range-based localization always has two phases to go: ranging and position computation. In the first phase it utilizes some ranging methods such as TOA (Time of Arrival) [13], TDOA (Time Difference of Arrival) [14], AOA (Angle of Arrival) [15] and RSSI (received signal strength indicator) [17] to obtain the distance between two nodes. With the coordinate information of the reference node attached with RSSI, the blind node (also refers to the mobile node or target) can calculate its own coordinate by using some mathematical methods, such as Trilateration, Triangulation, and Maximum likelihood estimation.

In TOA based localization system, it requires extra hardware to guarantee the synchronization between transmitting equipment and receiving equipment; otherwise, a small timing error may result in tremendous distance estimation error. In TDOA systems, it shares the same drawbacks with the TOA systems where they call for expensive hardware. Moreover, TDOA employs ultrasound ranging technique, which needs density deployment as the transmission distance of ultrasound is merely 20-30 feet. AOA localization system can be regarded as complementary technique for TOA and TDOA. It allows nodes to estimate the distance according to relative angles, which can be achieved by installing angle measuring equipment. For this point, it is also not advised to be used in large scale sensor networks. On the contrary, RSSI technique overcomes a majority of shortcomings mentioned above. It utilizes some signal propagation models, either from theoretical or empirical, to translate signal strength into distance. Thus it doesn't need additional hardware. However, this technique usually suffers from multi-path fading, noise interference, and irregular signal propagation, which has severely affected the accuracy of ranging estimate. Although existence of these disadvantage, we still can alleviate this suffering in some special methods, and we argue that only if proper measures are taken, the localization accuracy can be improved to meet the requirements of most application systems. We achieve this by employing regular deployment of node, region partition and localization refinement.

In RSSI localization systems, distance estimation between transmitter and receiver by using received signal strength based-on some signal propagation model should be accomplished previously. The widely used propagation model is *log-normal shadowing model* expressed as:

$$P_r(d)[dBm] = P_r(d_0)[dBm] - 10n\log_{10}(d/d_0) + x_\sigma[dBm] \qquad (1)$$

where $d$ is the distance between transmitter and receiver, $d_0$ denotes reference distance, $P_r(d)$ denotes the received power, $P_r(d_0)$ denotes the received power of the point with a reference distance $d_0$, $n$ denotes exponential attenuation factor to distance which is related to environment, and $x_\sigma$ presents Gaussian random variable which reflect the change of power when distance is fixed.

We use the simplified shadowing model:

$$P_r(d)[dBm] = P_r(d_0)[dBm] - 10n\lg(d/d_0) \qquad (2)$$

Usually, we select $d_0$ as 1 meter, so we have:

$$RSS[dBm] = P_r(d)[dBm] = A - 10n\lg d \qquad (3)$$

where $A$ is the received signal power of receiver from a transmitter one meter away.

Our system uses the CC2430 chip, which is system-on-chip solution for 2.4GHz IEEE 802.15.4/Zigbee with the characteristics of low-power and low date rate (up to 250kbps). CC2430 has a build-in register called $RSSI\_VAL$ for storing RSSI value, and the power value on RF pin is:

$$RSS = RSSI\_VAL + RSSI\_OFFSET [dBm] \quad (4)$$

Empirically, the $RSSI\_OFFSET$ can be assigned -45dBm. Up to now, we can estimate the distance between the transmitter and receiver.

This paper is partially based on our previous work [2], and extends to propose a reference model for location-based services (LBS) when using this localization algorithm.

## 3 Localization scheme

For many systems we hope to get location information to provide more intelligent services. For example, a mobile advertising business which needs to know the position of people for pushing services (discount shopping news), to find a nearest coffee shop according to current position, to track somebody in the stadium. Some systems have high precision positioning requirements, because the location accuracy directly impacts the performance of entire application system. For example, cargo tracking in large warehouse, wharf cargo scheduling, cargo location on crane tower.

In this section we present a localization method, which consists of two phases: region partition and localization refinement. The basic idea of this method is that we first split the target region into small grids by deploying sensor nodes regularly, and the nodes reside on the vertex of the grid. The grid in which the blind nodes current located can be easily determined by comparing their RSSI values. The shorter the distance, the larger the RSSI value is, and vice versa. In this determined grid we then refine the position coordinates to meet the accuracy requirement by employing a trick algorithm.

### 3.1 Region partition

To simplify this problem we consider dividing the target region into a number of grids (square or rectangle). Firstly we setup the coordinate XOY for a target region, and then draw lines which parallel to the $x$ and $y$ axis respectively. Actually the distance between any two lines are not constrained as a constant number; that's to say, you can set the distance between two neighbor lines to what you want according to the accuracy requirement of your system. Usually the distance can be 5 to 10 meters approximately, which is not a restriction of using our algorithm. If you want to improve the system precision, the distance between lines can be set as 5 meters or less, and if the precision requirement of your system is not so high, the distance can be set up to 50 meters or even 100 meters as long as the distance of this two nodes which should not exceed the radio communication radius. In the trial system, the distance between beacon nodes is set up to 4 meters, which will be described in next section. Its reason comes from our experimental result, which is illustrated in Figure 1 and Figure 2, showing that with the ranging distance increasing, the error of distance estimation becomes bigger and it is difficult to describe the variation of error through an equation. The figures together show the ranging results and their errors we have tested at the indoor and outdoor environments. In Figure 1, it is obvious that the calculated distances in outdoor environment are more concentrated and close to the real distance, while in indoor environments they are more scattered and there exists many irregular nodes with the distance increasing. Because in the indoor environment, wireless signal is more easily suffering from multipath interference, reflection, refraction, obstruction interference and so on. On the contrary, the interference of outdoor environment is much smaller. In consequence the shorter real distance is, the more accurate distance we will get. Figure 2 shows that the average error of different real distance in outdoor and indoor environment, from the results we can see when the real distance is less than four or five meters, the result obtains an acceptable ranging error. Thus in the localization system introduced in this paper we set the distance between any two reference nodes to four meters.

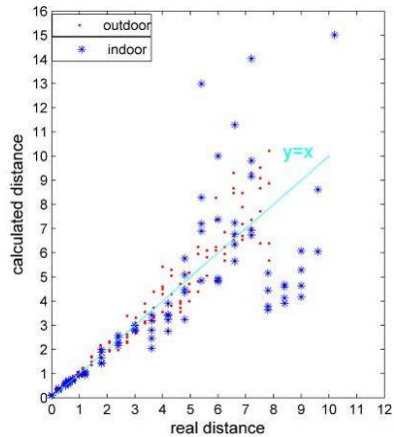

**Fig. 1** Real distance and calculated distance in outdoor and indoor environments

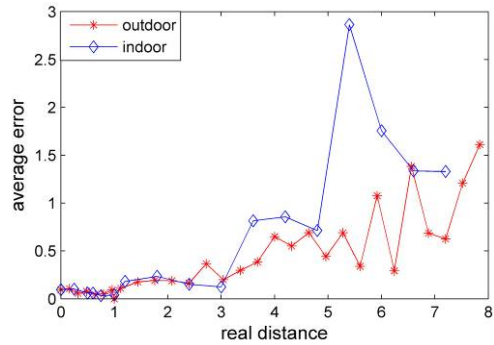

**Fig. 2** Average calculated error at different distances in outdoor and indoor environments

Next, we discuss which grid the blind node belongs to. Assume that the RSSI value can correctly reflect the distance indicating a blind node is near or far from a reference node. When the blind node moves into a grid, it broadcasts a message. Any reference nodes that receive this message can extract a RSSI value from it. After a short-time accumulation of RSSI values followed by average value calculation, a RSSI value is generated and returned to the blind node. For a blind node, it may receive many RSSI values from its surroundings via broadcast. However, we just pick out the maximum four values with their corresponding coordinators. The interactions between blind node and beacon nodes are described in Figure 3, in which the beacon nodes accumulate RSSI values 8 times and return the average to blind node. In some systems it may carry on more times of accumulations, but this will affect the timeliness because two RSSI accumulations need a time spacing interval, and less RSSI accumulations will make the system easier subject to environmental impact. Therefore, selecting a good accumulation number is also particularly important to the system performance.

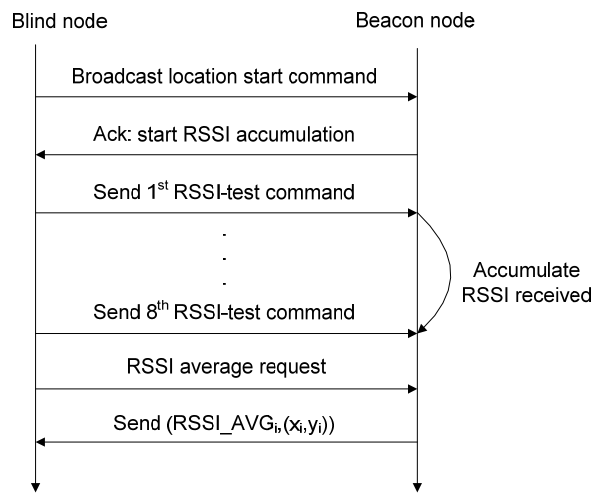

**Fig. 3** Interaction between blind node and beacon nodes to obtain RSSI values

From the above discussion, theoretically we can easily infer that the coordinate set can form a rectangle, in which the blind node resides. However, in practice the coordinate set may not be able to construct a rectangle when blind node moves from one grid to another or due to long period of blind node paging. So firstly we should examine whether the coordinate set can form a rectangle. If so, we can conclude that blind node is located in this rectangle. Otherwise, the coordinate set may not form a rectangle. Figure 4 depicts this circumstance, and Figure 4(a) shows the case in which the blind node moves from one grid to another, and Figure 4(b) shows the case in which blind node approaches to a reference node. The common characteristic of these two cases is that the four reference nodes (i.e. $A$, $B$, $C$, $D$) from which blind node receives four maximum RSSI values cannot form a rectangle. For the former case, we divide these four points into two groups, in each of which the difference values of two RSSI is minimal, for instance $(A, B)$ group and $(C, D)$ group. Then we calculate the x-axis value and y-axis value respectively, and the two parts make up the position coordinate of blind node. For the latter case, by using the maximum RSSI we can calculate the distance $r$ to one reference node, such as $A$ in Figure 4(b). We can say that the blind node is in the circle with $A$ as its center and radius is $r$. But how to calculate its direction, that's to say, how can we know the blind node reside in which grid now? We utilize some variables to record the former moment in which grid the blind node belongs to, and employ them to infer the current direction information. Till now, we can determine which grid blind node locates currently, and then we can constrain the localization error within the bound of half length of the grid edge.

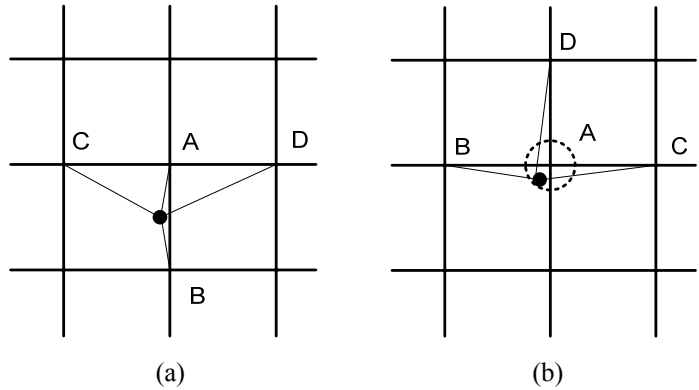

**Fig. 4** Cases where coordinate set cannot form a rectangle

### 3.2 Range accuracy improvement

In RSSI-based localization systems, the accuracy of the final location largely depends on the range. Therefore, in order to obtain higher accuracy, we need to try various possible methods to improve the range accuracy. In equation (3) we get the relationship of RSSI and distance, where the parameter $n$ means the exponential attenuation factor to distance. In the indoor environment, the external factors deeply affects range value; even in the same location of a node, the value of $n$ may be variable with different circumstances. In order to maintain a certain location accuracy, adjusting the value of $n$ is necessary. Here a method is designed to adjust the range value between blind node and beacon, which depends on RSSI and real distance of beacons.

As shown in Figure 5, the real distance between blind node $M$ and beacon node $B_1$ is $d_m$, the calculated distance is $d'_m$. The real distance between beacon nodes $B_1$ and $B_2$ is $d$, while the calculated distance is $d'$.

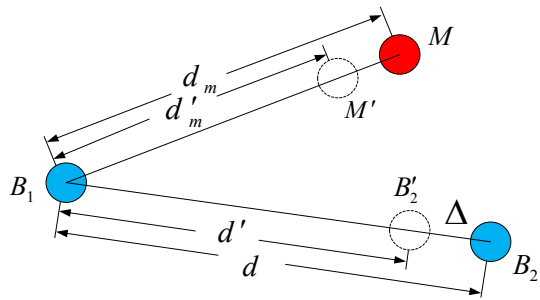

**Fig. 5** Range adjustment between nodes

From the formula (3), we get:

$$\begin{cases} RSSI = A - 10n \lg d \\ RSSI = A - 10n' \lg d' \end{cases}$$

Therefore, we have:

$$n = n' \times \log_d d' \qquad (5)$$

The initial value of path loss factor is specified as $n'$, and the real distance $d$ can be calculated according to the coordinator of two beacon nodes. $d'$ is the distance calculated last time. Thus, after several calculations by using this formula, we obtain the value $n$ which can exactly reflect the environment condition and using it we can calculate more precise distance between blind and beacon nodes.

### 3.3 Localization refinement

Although a rough position of blind node can be obtained as mentioned above, and the result may satisfy the requirement for the applications which only needs approximate location information or the systems which already possess a high density of reference nodes. However, by increasing the node density to gain a high precision is not a good choice which needs a number of reference nodes, which has become an obstacle for large-scale applications. To overcome this problem, we design a fine-grain localization method for blind node in a grid, as depicted in Figure 6, by using a compact algorithm which can easily implement two-dimensional plane localization with a regular deployment of reference nodes.

This algorithm requires that the reference node must be deployed along with the axis (x-axis and y-axis). In Figure 6, by exploring $\triangle AMB$ and $\triangle CMD$, we have a set of equations:

$$\begin{cases} d_1^2 - (x - x_1)^2 = d_2^2 - (x_2 - x)^2 \\ d_4^2 - (x - x_1)^2 = d_3^2 - (x_2 - x)^2 \end{cases} \qquad (6)$$

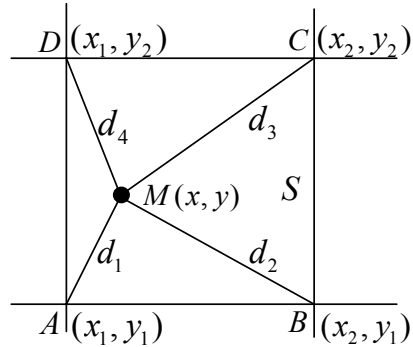

**Fig. 6** Position calculation of blind node in a grid

In practice, usually these equations has no solutions as the ranging errors are different for the four nodes. However, by using each of these equations we can calculate an *x*-coordinate value. The details are described as follows: 1) to calculate two *x*-coordinate values respectively by using equations (6), and 2) to calculate the average of these two values as the final *x*-coordinate value for blind node, which is:

$$x = \frac{1}{2} \cdot [x_1 + x_2 - \frac{(d_1^2 + d_4^2) - (d_2^2 + d_3^2)}{2(x_1 - x_2)}] \qquad (7)$$

Using the same method, we can easily obtain the *y*-coordinate value:

$$y = \frac{1}{2} \cdot [y_1 + y_2 - \frac{(d_1^2 + d_2^2) - (d_3^2 + d_4^2)}{2(y_1 - y_2)}] \qquad (8)$$

Therefore, the final coordinate of blind node consists of (7) and (8).

The overall flow chart of localization computation in blind node is shown in Figure 7. For the blind node as well as beacon nodes, the localization process pseudo-code is shown below.

**Algorithm 1** Blind node localization process algorithm

    Broadcast location start command.
    Wait for Ack;
    **IF** receive Ack **Then**
      **For** accumulate times=1:8
      **Do**
        Send RSSI-test command.
        Wait 20 ms.
      Send RSSI average value request command.
      Temporarily store each beacon nodes' ($RSSI_i$, ($x_i$, $y_i$)) ($1 \leq i \leq N$)
      **IF** ($N \geq 4$) **Then**
        Select up to four groups.
        **IF** four groups can form rectangle **Then**
          Use formula (7) (8) to calculate (x, y).
      **Else**
        **IF** in four groups exist a large RSSI **Then**
          Use last location to determine the direction $\theta$.
          Use RSSI value to get distance $d$.
        **Else**
          Calculate (x, y) in two groups by dividing the four groups.
    **Else**
      Return.

For beacon nodes the process is as follows:

**Algorithm 2** beacon node localization process algorithm

    **Do**
      Wait for RSSI accumulation start command.
      Send Ack to blind node to start RSSI test.
      **For** each received RSSI-test from blink node **Do**
        **IF** receive RSSI average request **Then**
          Calculate the average value of accumulated RSSI and return to blind node.
        **Else**
          Accumulate RSSI value.
    **While** the application is running.

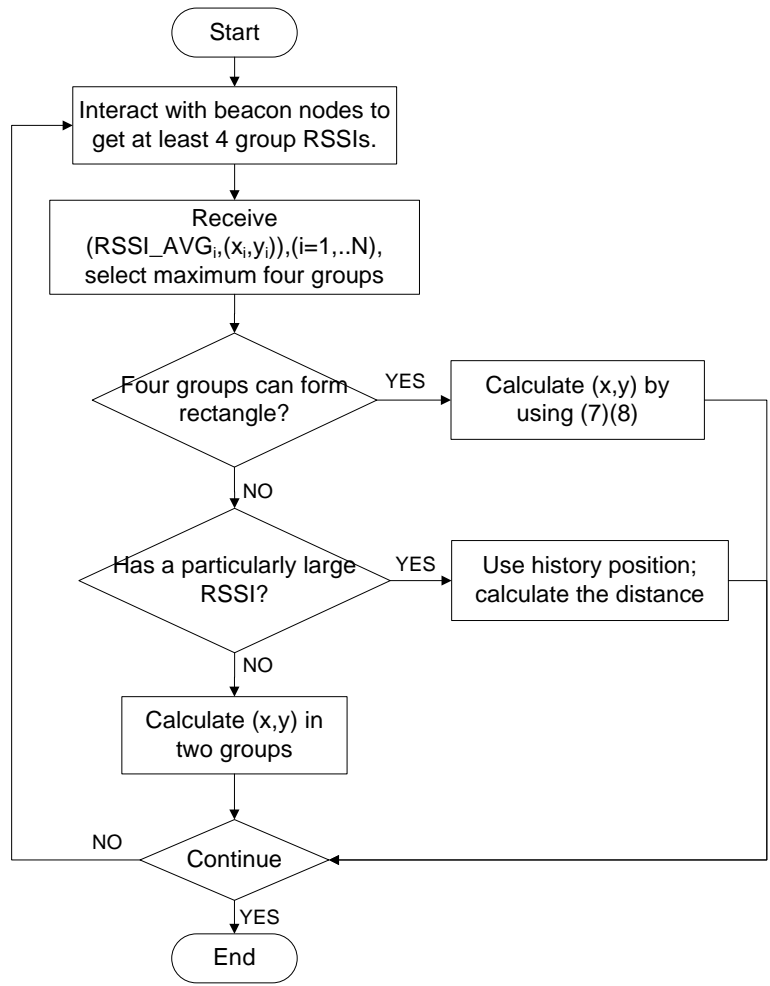

**Fig. 7** Blind node localization process flow chart

3.4 Localization algorithm for LBS

For a localization algorithm, on one side it probably provides interface or simple service for other modules, for instance, identification for object, location based routing algorithms. On the other side, however, it mainly provides with more intelligent location based services. The IOT has afforded a perfect opportunity and platform to carry out LBS, as it integrates mobile terminal, localization technology, communication network and content provider together, providing a new way for location information transmission and usage.

In this paper we propose a location based services reference model as depicted in Figure 8, which can well make use of the algorithm we suggested above. This model can be divided into three layers: the bottom is data acquisition layer, in which relevant sensor raw data and location interrelated information are collected. The middle is for data process and transmission layer, which is responsible for data abstraction of low layer and formatting context information for other modules or upper layer. Localization module can use a concrete localization algorithms to calculate the location of an entity, and provides the interface for the upper applications, generally. This localization module in our localization system uses the above-mentioned localization algorithm. The upper layer is for application services, which invokes the lower interface or purchases the data from content providers to provide intelligent services, simultaneously collecting revised feedback information back to context module. Through this constant iteration, the system will become more optimized and more intelligent. In the next section we will introduce a demonstration system based on this model and apply the proposed localization algorithm. The whole system is based on the background of digital home in IOT, and realizes the intelligence and better service after employing location information.

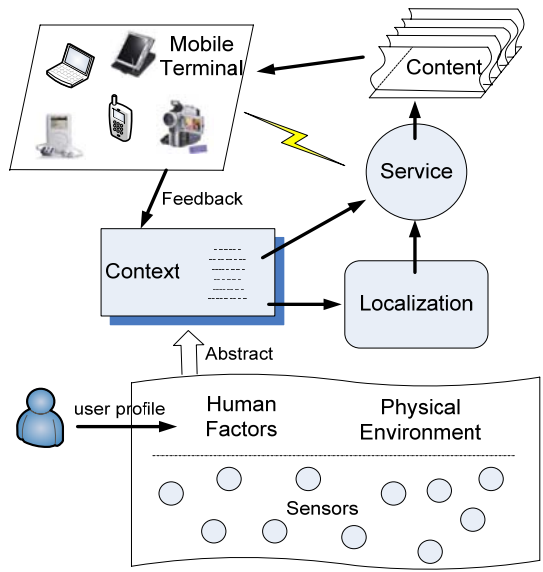

**Fig. 8** A reference model for location-based services

## 4 Performance evaluation

### 4.1 Localization accuracy

The proposed scheme is implemented in a realistic environment. The chip we used is cc2430 produced by TI Corporation, and it is a system-on-chip solution for short distance wireless communication applications. It has integrated RF components, which provide RSSI information with ability of low energy consumption, thus it can construct the proposed localization application easily. Also, we run a Zigbee stack named z-stack (version: 2.4.3) made by TI Corporation on this chip platform. Although a localization system has existed and completed by them using the CC2431 chip, the function of which almost equivalent to the CC2430 except for owing a localization engine. The localization error is within 3 meters in our test environment. However, in this system we found that when the blind node moves near to the reference nodes, localization errors usually larger than in the other region.

To illustrate the performance of our system, we compare it with the CC2431 localization system. Figure 9(a) shows our outdoor test environment and Figure 9(b) is the real node used in trial system. Both of these two systems are deployed at the same place during nearly the same period time, with the same spacing distance (four meters) of reference nodes. Also we try our best to make the other environment related differences of these two systems are as small as possible at run time. We capture 625 data in 8m*8m square area. Figure 10 shows the 3D localization results and errors comparison in different points. The z-axis shows the error value, which represents the distance between calculated coordinate (x', y') and real coordinate (x, y). On the whole, our system in Figure 10(b) captures a more accurate localization results compared with CC2431 system in Figure 10(a), on both the edge of two reference nodes and the area near to them.

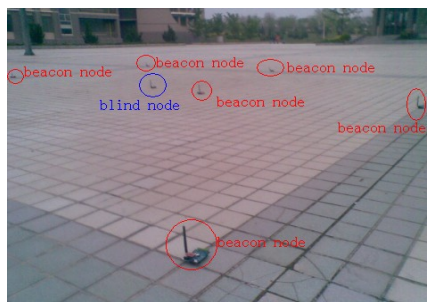
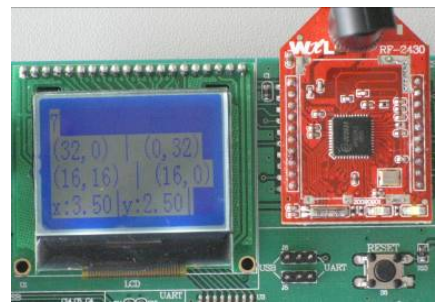

(a) (b)

**Fig. 9** Trial environment and sensor node

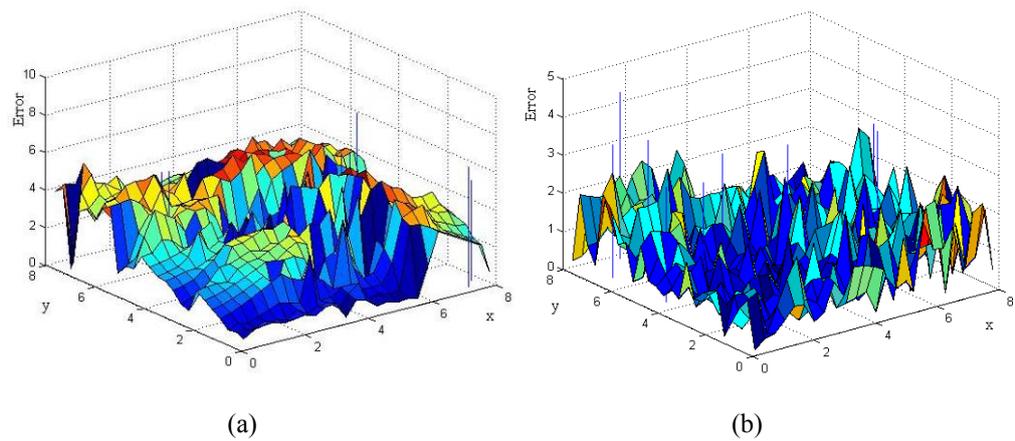

**Fig. 10** Error values in different areas: (a) Error distribution of TI CC2431 localization system in the test environment. (b) Error distribution when using the proposed algorithm in the same environment

According to statistical analysis of 625 data, we get the proportion distribution in different error intervals, which is presented in Figure 11. From it we can see that the error distributions of our system are concentrated within 2.5 meters, and the error values less than 1.5 meters take nearly 80% of the whole system. While the localization errors of CC2431 system below 3 meters practically account for 50.4%, and the error values more than 3 meters take 49.6% in the experiment.

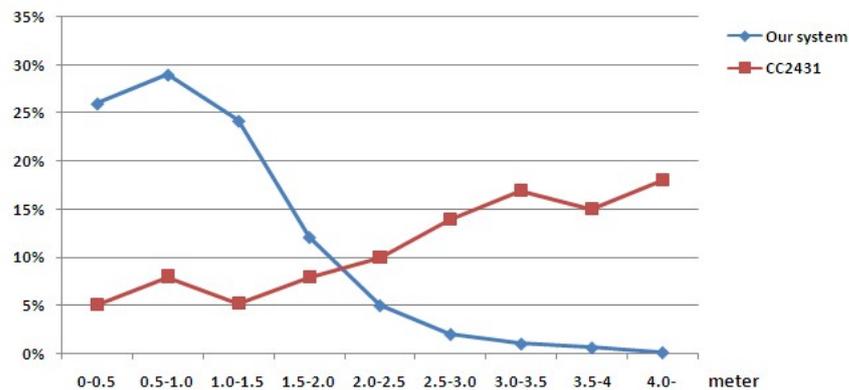

**Fig. 11** Error distribution of these two systems

### 4.2 Real system

We integrate this algorithm into the location based services reference model as mentioned before. We use this approach to locate mobile entity such as mobile people, mobile terminal, and mobile sensor node. In our demonstration system we first locate some nodes with sensors such as temperature and humidity sensors, light sensor, gas sensor, and smoke sensor. When the temperature have changed in a way or detect some flammable gas or smoke, it will trigger alarm and inform the master. Also a person with a mobile node goes around in the room; the real time location information will transfer to controller via gateway. We also use human infrared sensor to assist to detect some person who tries to intrude the room.

For evaluating the proposed system, we design a digital home system with various sensors, including above mentioned sensors with other complementary nodes such as window curtain controller, appliance switch control node, lamp brightness adjustment node, acceleration sensor node, tilt sensor node, Infrared radio and TV controller node, Bluetooth data and voice transmission nodes. Using the proposed scheme combining sensing parameters by those sensors, we can easily build some interesting and intelligent applications, such as to know status of the appliances in digital home system and to control them. For example, our curtain will automatic open or close in a location or angle according to the data from photo sensor. Meanwhile, the designed digital home system can be used to manage university's laboratories. We can set up the scene pattern to detect when we leave our lab or whether someone visits the lab. If someone enters our room, the system will be triggered to record the visiting time, to track the person and to notify the manager by short message. Figure 12 shows partial sensors that we designed for digital home to assist our localization system, which includes 2 cameras, one temperature and humidity, one tilt, one personal infra, infra convertor to control radio and DVD player, two lamp brightness adjustments, one terminal, one smart socket and one intelligent gate way.

All of these applications are based on the three layers reference model proposed in this paper, which helps us easily make the system more intelligent and more interesting.

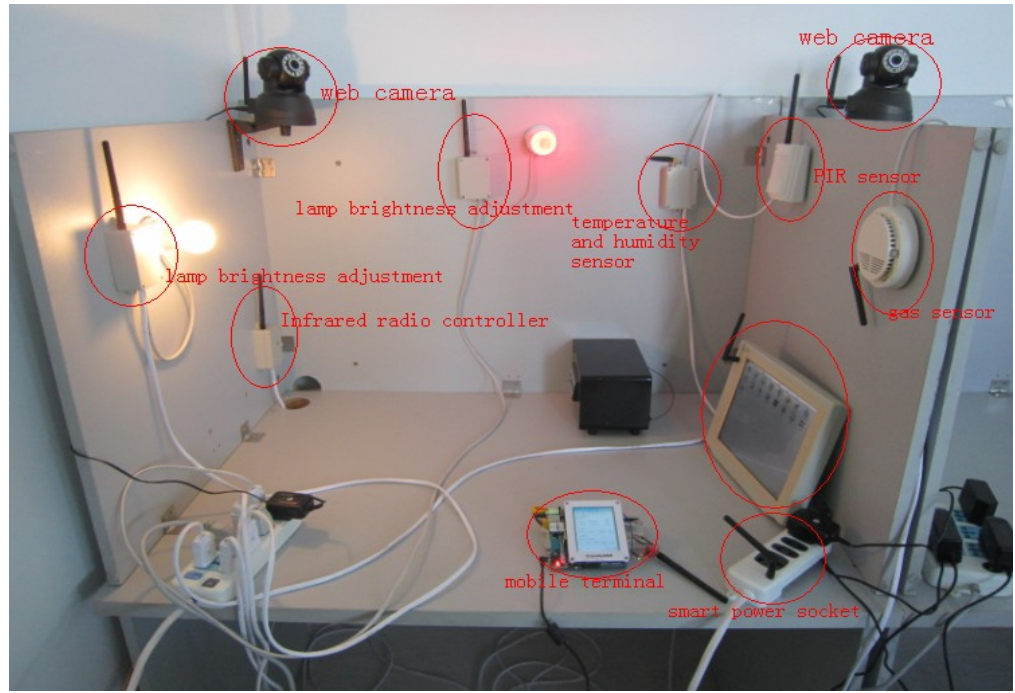

**Fig. 12** Some devices used in the system

## 5 Conclusions

Localization is a very critical issue in the IOT, which is also a very challenging task facing WSN. As the location information is regarded as basic information in many applications, great efforts have been made by many researchers and a variant of algorithms have been proposed. In this paper we mainly focus on the usability and availability in realistic environment and application systems. It is a good choice to establish a location-based service system for IOT. Compared with the traditional algorithm, the proposed scheme is lightweight and succinct. The experimental results show that the localization performance based on the proposed algorithm is better as compared with the CC2431 system under the same test environment. In addition, a location-based services reference model is designed. Based on this model, a trial system is designed to illustrate that the proposed scheme can be well applied to IOT systems.

The proposed scheme still have room to be improved. For example, if the blind node moves with a high speed, the localization precision will descend. Secondly, how to maintain the almost same performance when the distance between two reference nodes is enlarged. All these problems also exist for most of the other localization schemes in WSNs. More ingenious algorithms are expected to be designed to solve these problems. All these efforts will greatly accelerate the development of the IOT technique to be applied into real application systems.

### Acknowledgements

This work was partially supported by the Natural Science Foundation of China under Grant No. 60903153, the Fundamental Research Funds for Central Universities (DUT10ZD110), the SRF for ROCS, SEM, and DUT Graduate School (JP201006).